\documentclass[onecolumn,showpacs,prl,superscriptaddress]{elsarticle}
\usepackage{graphicx}
\usepackage{color}

\usepackage{etoolbox}
\apptocmd{\thebibliography}{\raggedright}{}{}

\begin{document}

\title{Modelling the downhill of the Sars-Cov-2 in Italy and a universal forecast of the epidemic in the world}

\author[rvt]{Gabriele Martelloni}
\ead{gabriele.martelloni@gmail.com}
\author[rvt]{Gianluca Martelloni}
\ead{gianluca.martelloni@unifi.it}
\address[rvt]{University of Florence Ugo Schiff Department of Chemistry- INSTM - via della Lastruccia 3-13, I - 50019 Sesto Fiorentino, Firenze, Italy.}

\begin{abstract}
In a previous article \cite{MM} we have described the temporal evolution of the Sars-Cov-2 in Italy in the time window  February 24-April 1. As we can see in \cite{MM} a generalized logistic equation captures both the peaks of the total infected and the deaths. In this article our goal is to study the missing peak, i.e. the currently infected one (or total currently positive). After the April 7 the large increase in the number of swabs meant that the logistical behavior of the infected curve no longer worked. So we decided to generalize the model, introducing new parameters. Moreover, we adopt a similar approach used in \cite{MM} (for the estimation of deaths) in order to evaluate the recoveries. In this way, introducing a simple conservation law, we define a model with 4 populations: total infected, currently positives, recoveries and deaths. Therefore, we propose an alternative method to a classical SIRD model for the evaluation of the Sars-Cov-2 epidemic. However, the method is general and thus applicable to other diseases. Finally we study the behavior of the ratio infected over swabs for Italy, Germany and USA, and we show as studying this parameter we recover the generalized Logistic model used in \cite{MM} for these three countries. We think that this trend could be useful for a future epidemic of this coronavirus.

\end{abstract}


\maketitle
\section{Introduction}
We briefly review the historical evolution of the Sars-Cov-2 in the Earth. In early December the Sars-Cov-2 appeared in Wuhan, China.\\
The disease caused by the new Coronavirus has a name: "COVID-19" (where "CO" stands for corona, "VI" for virus, "D" for disease and "19" indicates the year in which it occurred). The Oms Director-General Tedros Adhanom Ghebreyesus announced it on February 11, 2020, during the extraordinary press conference dedicated to the virus.\\
The appearance of new pathogenic viruses for humans, previously circulating only in the animal world, is a widely known phenomenon (called spill over) and it is thought that it may also be at the basis of the origin of the new coronavirus (SARS- CoV-2). The scientific community is currently trying to identify the source of the infection.\\
On December 31, 2019, the Municipal Health Commission of Wuhan (China) reported to the Oms a cluster of cases of pneumonia of unknown etiology in the city of Wuhan, in the Chinese province of Hubei.\\
On January 9, 2020, the Chinese Center for Disease Prevention and Control (CDC) reported that a new coronavirus (initially called 2019- nCoV and now called SARS-CoV-2) has been identified as the causative agent and has been rendered publishes the genomic sequence.\\
Oms on March 11, 2020 declared that COVID-19 can be defined as a pandemic.\\
After notification of the epidemic by China, Italy immediately recommended postponing unnecessary flights to Wuhan and, subsequently, with the spread of the epidemic, to all of China.
Consequently, the latter has canceled all flights from Wuhan.\\
This disease does not save Italy that has become a protected area with the DPCM signed on the evening of 9 March by the Prime Minister, Giuseppe Conte, who has extended the restrictive measures already applied for Lombardy and the 14 northern provinces most affected by the coronavirus infection to the whole national territory. The new action comes into force on March 10 and will take effect until April 3. Among the main innovations: it limits the movement of people, blocks sporting events, suspends teaching activities in schools and universities throughout the country until April 3.\\
With the new ordinance of 22 March 2020 issued by the Minister of Health and the Minister of the Interior, from 22 March people are prohibited from moving with public or private trasportation in a municipality other than that in which they are located, except for proven work needs, absolute urgency or for health reasons.\cite{salute.gov.it}\\
Many growth models have been very recently applied to study the evolution of the Covid-19 infection\cite{LBAGZ, CLG, Casto1, Fenga, FP, AG, Betal, Casto2}. 
In \cite{MM} we tried to analyze the time evolution of the Sars-Cov-2 in Italy, using a Logistic model\cite{Logistic} at the beginning of the study and after with a generalization of that model. \\
The Logistic behaviour assumes that growth stops when maximum sustainable population density is reached through the carrying capacity K that depends on the environmental conditions. For example the ordinances of the Prime Minister G.Conte, the people's hygiene habits are encoded in the carrying capacity K.\\
We observe as the generalized model of \cite{MM} works very well until the April 7. After this date the large increase in the number of swabs meant that the logistical behavior of the infected curve no longer worked. At first in Italy, pharyngeal swabs were initially made only on seriously ill people. This choice gave us the possibility to have a sample of the infected that we can describe with a single population model, after April 7 it becomes impossible. So we decided to use a different model to describe the new trend of the data and try to give different scenarios of the descent phase of the virus in Italy, in the time window  February 24-May 5. In \cite{MM} we described two different peaks, the peak of the infected and the deaths one. In this paper we analyze the peak of the currently infected and the downhill of the propagation of the Sars-Cov-2. To do this we define a new model similar to a SIRD (see for example \cite{FP} ), but without the population of supsceptibles, because there are no criteria on defining the susceptible ones. We consider three couple differential equations for Infected I(t), Deaths D(t) and Recovery R(t) with the following conservation law 
\begin{equation}
P(t)=I(t)-R(t)-D(t),
\end{equation}
where $P(t)$ represents the currently infected (or positive).\\
In the last part of this article we observe as the following ratio (infected $I(t_{i})$ over swabs $S(t_{i})$)
\begin{equation}
I_{\mbox{norm}}(t_{i})=\frac{I(t_{i})}{S(t_{i})},
\end{equation}
is the most important parameter to describe the evolution of the Sars-Cov-2. Indeed, we can describe the trend of this quantity only with a generalized Logistic model with 4 parameters even with data after April 7. This behavior suggest us to use this model for a future epidemic of this virus. If we will able to perform a greater and constant number of swabs everyday, using this model, we may have better control over the contagion curve, and consequently over the number of deaths.

\section{The new model and the description of the dowhill}
Our idea is to use a model that adapts to the data of the problem. We explain better. Let's consider the following data:
\begin{itemize}
\item 1) daily data of March 14: 3497 new infected, 11682 swabs;
\item 2) daily data of April 17: 3493 new infected, 65705 swabs;
\item 3) probable Case Fatality Rate (CFR) of the virus is rougly $1\%\rightarrow$ 2.3 million of infected at April 17, with 172434 detected;
\item 4) time delay between hospitalization and death $t_{d}\simeq 4$ days, parameter extrapolated also in \cite{MM};
\item 5) time delay between the onset of symptoms and healing $t_{r}\simeq 14-42$ days, a very oscillating parameter;
\item 6) two different LockDown (LD) data, March 10 and March 22, with different restriction.
\end{itemize}
Some comments about these data: the points 1) and 2) describe perfectly that the sample of infected is not clean; at the beginning of the contagion the swabs are performed only on the severe infected, after 1 month the number of swabs are increased of a factor 6 and consequently also the midly infected are detected. Point 3) tells us that there is probably an incredible number of asymptomatics as a source of severe infected, we have no control about it. Points 4), 5) indicate that while the death data is under control, the healed data are very oscillating in time. Finally the points 6) tells us that contribution of asymptomatics, portrayed in \cite{MM}, changes in time, indeed from April 6-7 (14-15 days after the second LD, i.e. an incubation time ) the generalized Logistic description fails. \\
After these considerations we have decided to couple the following equations:
\begin{eqnarray}
\frac{d I}{d t} &=& r_{0} I^{\alpha}+A t^{\beta} \qquad t\leq t_{0}, \\
\frac{d I}{d t} &=& r_{0} I^{\alpha}-\frac{r_{0} I^{\delta}}{K}+A t^{\beta}t^{-\gamma(t-t_{0})} \qquad t> t_{0},\\
\frac{d D}{d t} &=& K_{f} \frac{d I(t-t_{d})}{d t}, \\
\frac{d R}{d t} &=& f(t) \frac{d I(t-t_{r})}{d t},
\end{eqnarray}
with a conservation law
\begin{equation}
P(t)=I(t)-R(t)-D(t),
\end{equation}
where $P(t)$ represents the currently infected (or positive).\\
The parameters $r_{0}$ represents the rates of growth of epidemic, K is the carrying capacity for the classical logistic model, $\alpha$ is a constant in order to have a power low initial growth before LD, $\beta$ is the exponent of the second term of equation 1 that represents the influence of asymptomatic; $\delta$,a correction of the quadratic term of logistic, and $\gamma$ are the constant parameters considering the influence of the government measures\footnote{Remember that the measures of LD are not homogeneous in time, two different DPCM are present in the period March 10-May 4}, $K_{f}$ is a proportionality constant between deaths and total number of infected, while $t_{d}$ and $t_{r}$ are the delays of deaths and recoveries respect to infected respectively; the constant $A$ represents the contribution of asymptomatic people as introduced in \cite{MM} and finally $t_{0}$ is the time of LD start.\\
A brief consideration about the function $f(t)$: the great variability of $t_{r}$ suggest us that only the parameter $t_{r}$ is not sufficient to describe correctly the function R(t), so we decided to introduce a coefficient time dependent.\\
We present two different scenarios, in Fig.1 we consider a linear approximation $f(t)=a+bt$, while in Fig.2 we consider a quadratic approximation $f(t)=a+bt+ct^{2}$. This choice is not random. Indeed, considering the behaviour of the recovery time series in which a single recovery can heal with some delay in a window variable from few days to two months, the correct modeling could be a regressive linear function of type $R(t)=\sum_{i=1}^N a_{i}*I(t-t_{i})$  (eventually introducing also no-linear term in the series), but in this way we introduce many degree of freedom how many are the coefficient $a_{i}$ of the regressive function. Therefore, we consider an approximation using the two functions $f(t)$ considered above. \\
We desumed the following values for the principal parameters by means of 100 stochastic simulation using direct method Gillespie algorithm adapted to non-autonomous differential equations:
\begin{eqnarray}
r_{0} &=& 0.29\pm 0.02,\\
K &=& 1.85\ast 10^{5}\pm 1.1\ast 10^{4},\\
t_{0} &=& 30\pm 1,\\
\delta &=& 1.84\pm 0.01,\\
K_{f} &=& 0.142\pm 0.002,\\
t_{d} &=& 4\pm 1,\\
t_{r} &=& 8\pm 1 \quad \mbox{for linear approximation},\\
t_{r} &=& 12\pm 1 \quad \mbox{for quadratic approximation}.
\end{eqnarray}
\begin{figure}[hbt!]
\center\includegraphics[width=0.45\textwidth]{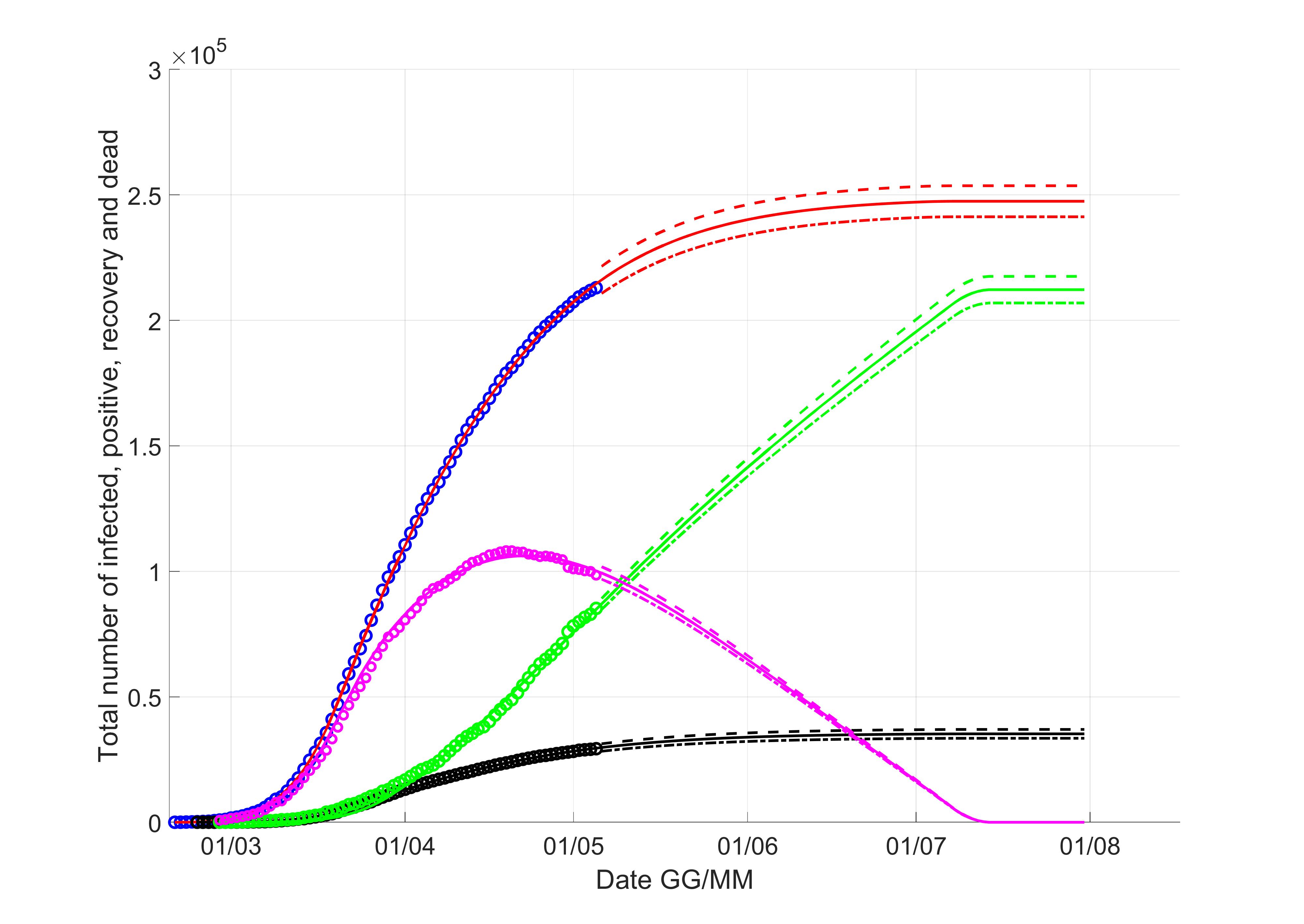}
\caption{The scenario with a linear growth for the recoveries: the black curve represents the deaths, the red one for the infected, the green one for the recovery and the pink one for the currently infected.} 
\label{Scenario1}
\end{figure}
\begin{figure}[hbt!]
\center\includegraphics[width=0.45\textwidth]{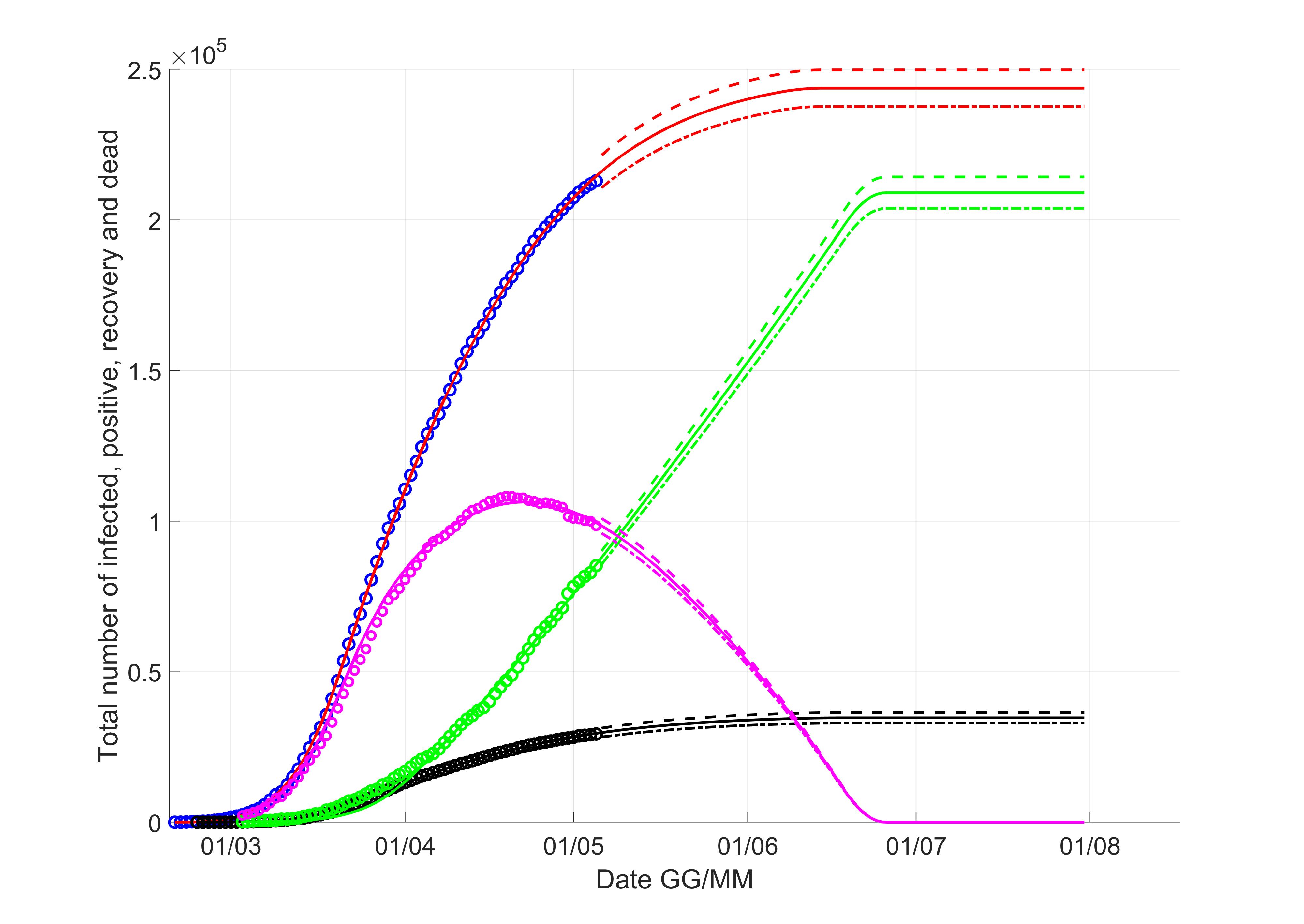}
\caption{The scenario with a quadratic growth for the recoveries: the black curve represents the deaths, the red one for the infected, the green one for the recovery and the pink one for the currently infected.}
\label{Scenario2}
\end{figure}
Some comments about these values: with respect to \cite{MM} we observe that the peak of the severe infected is correctly estimated, i.e. $t_{0}=24-26$ March and also the peak of the deaths, i.e. $t_{0}=28-30$ March; also the time delay $t_{d}$ remained the same; the same $t_{r}$ approaches the experimental lower limit in quadratic approximation. With respect to the Logistic model $r_{0}$ is increased while the coefficient $\delta$ drops from the value 2 to the value 1.84, i.e. we are considering different models.\\
In Fig. 1-2 we observe as the peak of currently infected is close to April 20 and finally we give us our prevision for a linear growth for the recoveries 
\begin{equation}
\nonumber
I(\mbox{end})=247471,
D(\mbox{end})=35235,
\end{equation}
close to July 10; for a quadratic growth we have
\begin{equation}
\nonumber
I(\mbox{end})=243766,
D(\mbox{end})=34682,
\end{equation}
close to June 20.\\
The estimated numbers $I(\mbox{end})$ and $D(\mbox{end})$ are very close, but it is not surprising: the eqs. (3) and (4) for total infected head the model, while $f(t)$ is present in eq. (6) that is only a proportionality equation. Obviously a linear approximation for f(t) leads to a slower recovery curve and therefore a small increase of infected.
\section{Recovering the Logistic behaviour}
Now we consider the following parameter:
\begin{equation}
I_{\mbox{norm}}(t_{i})=\frac{I(t_{i})}{S(t_{i})},
\end{equation}
that represents the number of infected normalized with the number of swabs $S(t_{i})$.\\
We study this quantity with generalized Logistic equation used in \cite{MM}:
\begin{equation}
\frac{d I_{\mbox{norm}}(t)}{d t}=r_{0} (I_{\mbox{norm}}(t)^{\alpha}-\frac{I_{\mbox{norm}}(t)^{2}}{K})+A, 
\end{equation}
where $\alpha$, $r_{0}$, $K$ and $A$ have the same meaning used in the previous section. Compared to the previous section we observe as studying the parameter $I_{\mbox{norm}}(t_{i})$ we can describe the contagion with a simple logistic equation and without the phenomenological terms introduced in eqs.(3)-(6). \\
In order to calibrate this model in the best way possible we use two algorithms, the first one based on simulated annealing \cite{SIMPSA} and the second one on optimized simplex \cite{SIMPLEX}. We evaluate the function error defined as
\begin{equation}
F(\mathbf{p})= 1/N \sum_{i=1}^N w_{i} (x_{i}-y_{i}(\mathbf{p}))^{2}
\end{equation}
where $x_{i}$ is the real data at day $i$, $y_{i}(\mathbf{p})$ is the correspondent output of the model depending of vector parameter $\mathbf{p}$ and $w_{i}$ is a generic weight that we can use or can be equal to one. For our purpose we adopt as weight the derivative of data or the data at time (day) $i$: the use of derivative allows to calibrate better on average the curve, while the use of the data as weight permit to calibrate better the data of the last part of the curve.
\begin{figure}[hbt!]
\center\includegraphics[width=0.45\textwidth]{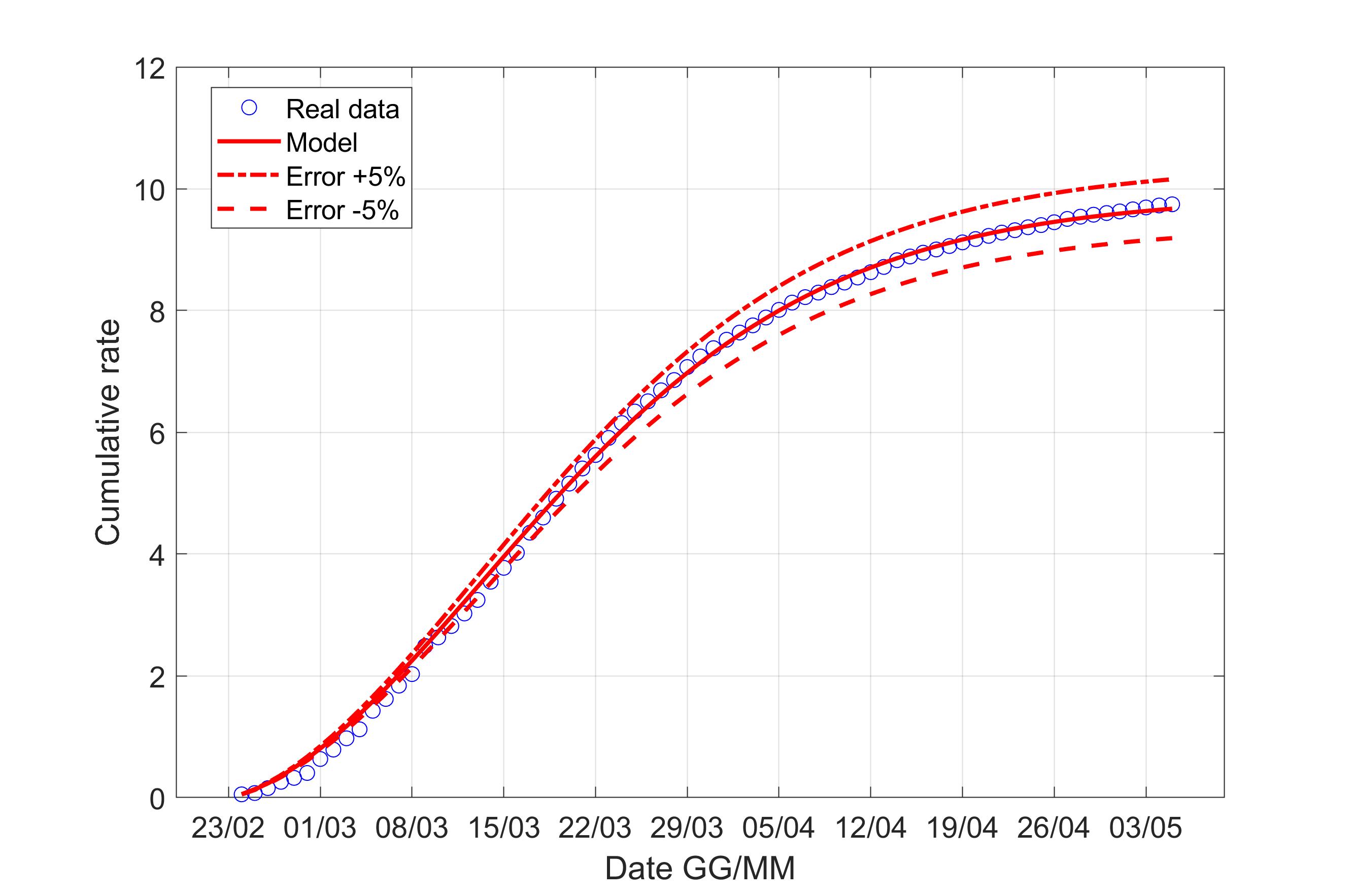}
\caption{The scenario of Italy with the derivative weight.} 
\label{Scenario3}
\end{figure}
\begin{figure}[hbt!]
\center\includegraphics[width=0.45\textwidth]{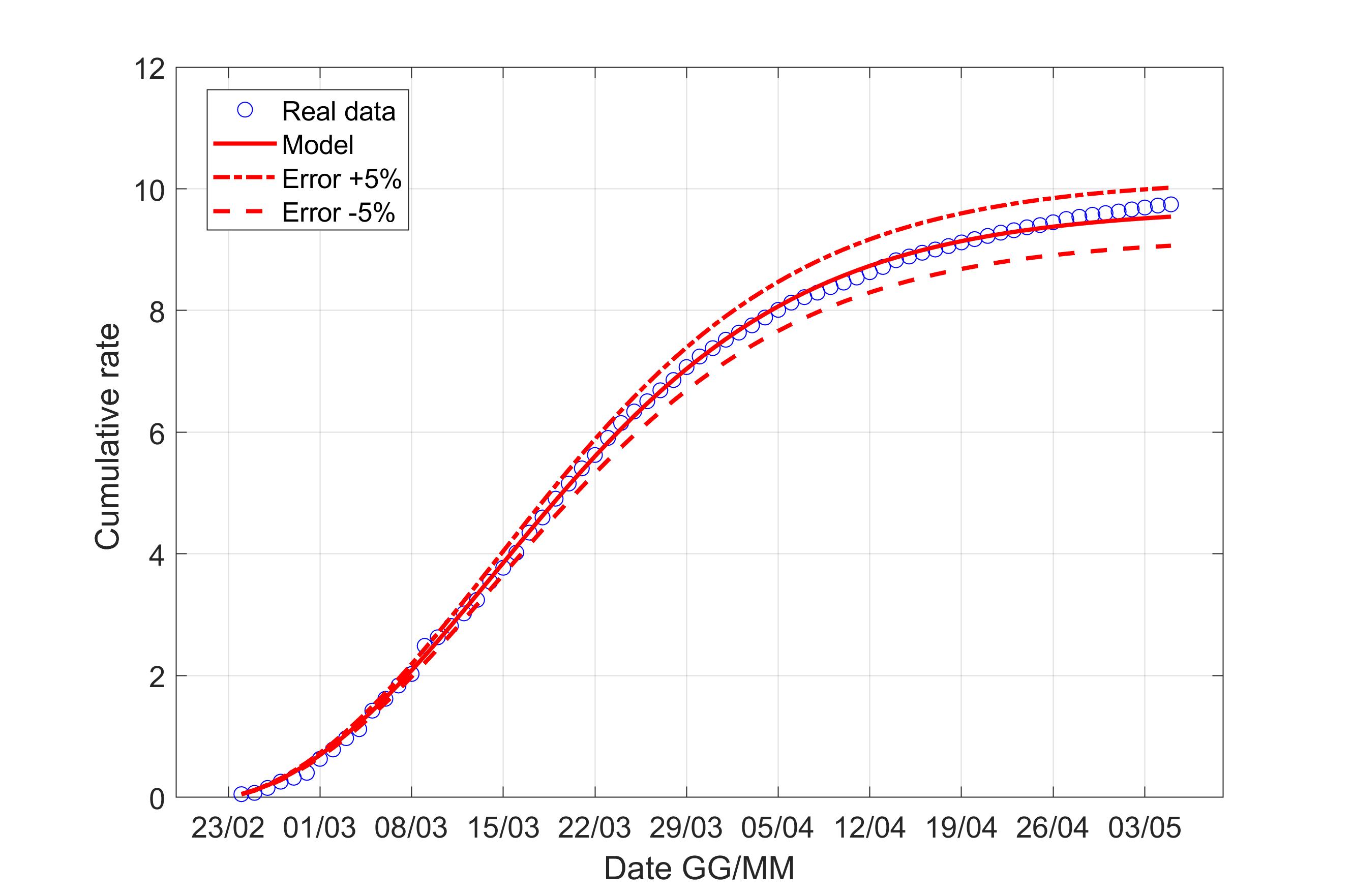}
\caption{The scenario of Italy with the data weight.}
\label{Scenario4}
\end{figure}
\begin{figure}[hbt!]
\center\includegraphics[width=0.45\textwidth]{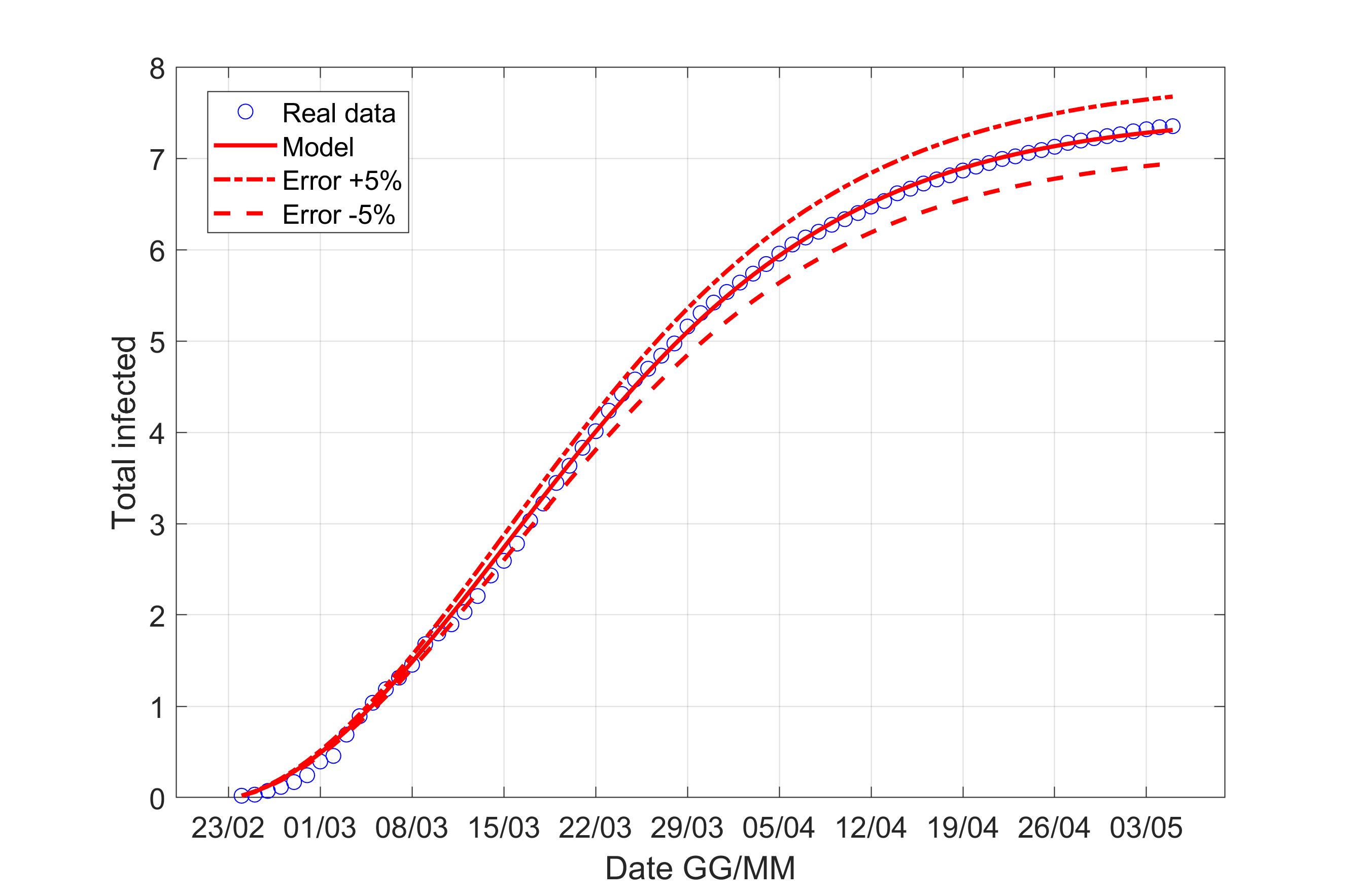}
\caption{The scenario of Italy minus Lombardia with the data weight.} 
\label{Scenario5}
\end{figure}
\begin{figure}[hbt!]
\center\includegraphics[width=0.45\textwidth]{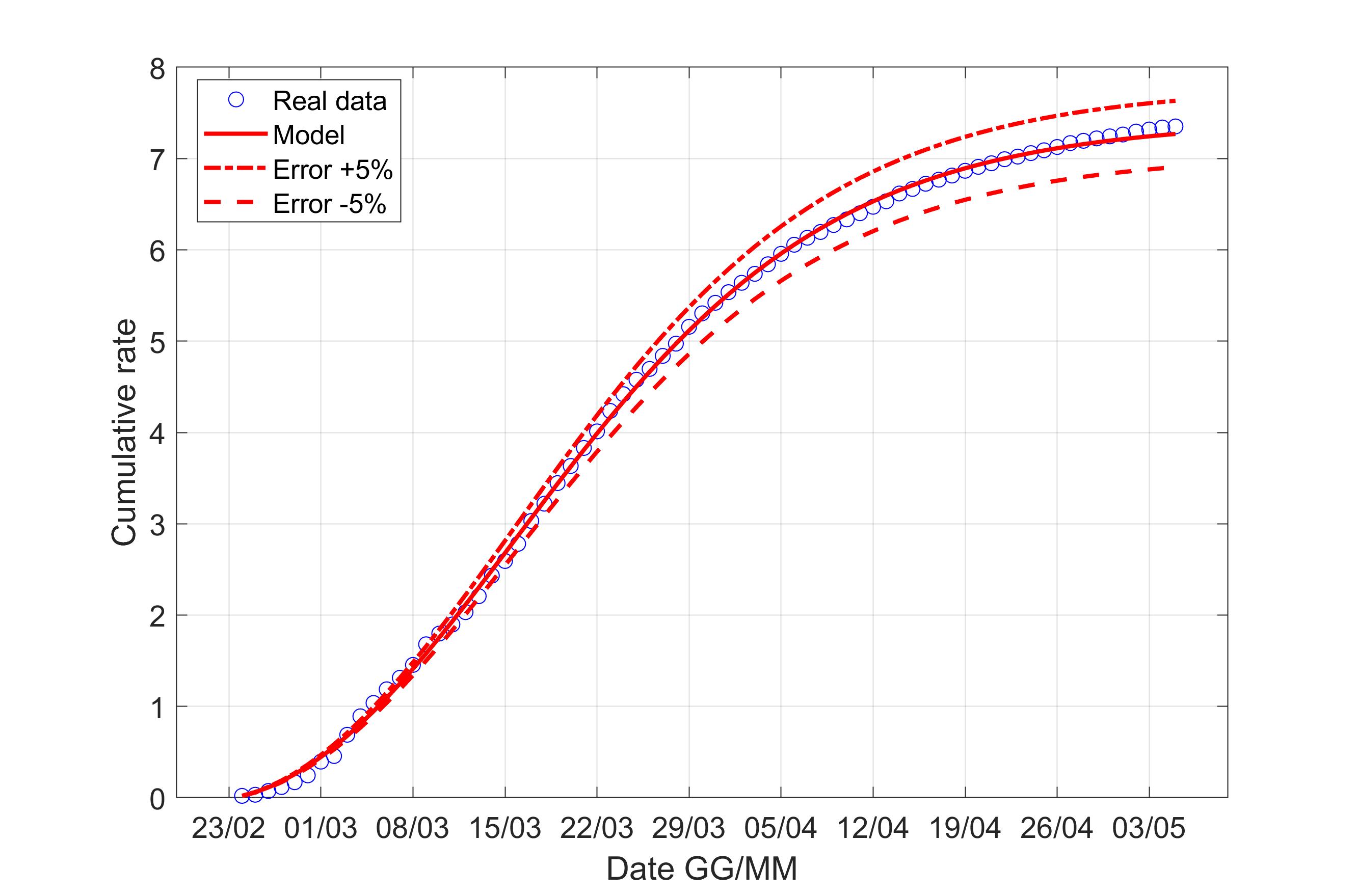}
\caption{The scenario of Italy minus Lombardia with the derivative weight.}
\label{Scenario6}
\end{figure}
In Fig. 3-4 we describe the situation of Italy at May 5, while Fig. 5-6 describe describe the data of Italy from which we have subtracted the data of Lombardia.\\
We give you the following values for the parameters of Fig. 3
\begin{eqnarray}
r_{0} &=& 0.175\pm 0.015,\\
K &=& 31.2\pm 0.6,\\
\alpha &=& 0.476\pm 0.08,\\
A &=& 0.010\pm 0.003,
\end{eqnarray}
for the parameters of Fig. 4
\begin{eqnarray}
r_{0} &=& 0.178\pm 0.015,\\
K &=& 38.5\pm 0.6,\\
\alpha &=& 0.398\pm 0.08,\\
A &=& 0.011\pm 0.003,
\end{eqnarray}
for the parameters of Fig. 5
\begin{eqnarray}
r_{0} &=& 0.143\pm 0.015,\\
K &=& 22.3\pm 0.4,\\
\alpha &=& 0.444\pm 0.08,\\
A &=& 0.011\pm 0.003
\end{eqnarray}
and finally for the parameters of Fig. 6
\begin{eqnarray}
r_{0} &=& 0.143\pm 0.015,\\
K &=& 20.2\pm 0.4,\\
\alpha &=& 0.485\pm 0.08,\\
A &=& 0.010\pm 0.003.
\end{eqnarray}
We observe as the quantity $I_{\mbox{norm}}(t)$ is probably the most important quantity studying the evolution of the virus! We explain better: the contribution of asymptomatic people is essentially the same in Lombardia and in the rest of Italy, while the coefficient $r_{0}$ is larger if we consider Italy compared to the scenario of Italy minus Lombardy; this consideration is extremely coherent with the data: the infected of Lombardia region represent the $37\%$ of all the italian infected. \\
Moreover the ratio infected over swabs is a very reliable parameter, we can describe correctly the italian situation only with 4 parameter and with a well-known model. We stress that in the future if a nation is ready to carry out a large and constant number of swabs every day, using this model, we can have a reliable forecast of the epidemic! 

\subsection{A comparison with Germany and USA}
We consider also the scenario represented by eq. (19) for Germany in Fig. 7 and for USA in Fig. 8. For Germany we study the time evolution of the Sars-Cov-2 in the time window March 8-May 11 and we obtain the following parameters
\begin{eqnarray}
r_{0} &=& 0.069\pm 0.008,\\
K &=& 8.6\pm 0.2,\\
\alpha &=& 0.61\pm 0.08,\\
A &=& 0.013\pm 0.003.
\end{eqnarray}
For USA we study the contagion in the time window March 10-May 11 and we have these values for the parameters
\begin{eqnarray}
r_{0} &=& 0.082\pm 0.008,\\
K &=& 29.9\pm 0.4,\\
\alpha &=& 0.71\pm 0.09,\\
A &=& 0.012\pm 0.003.
\end{eqnarray}
\begin{figure}[hbt!]
\center\includegraphics[width=0.45\textwidth]{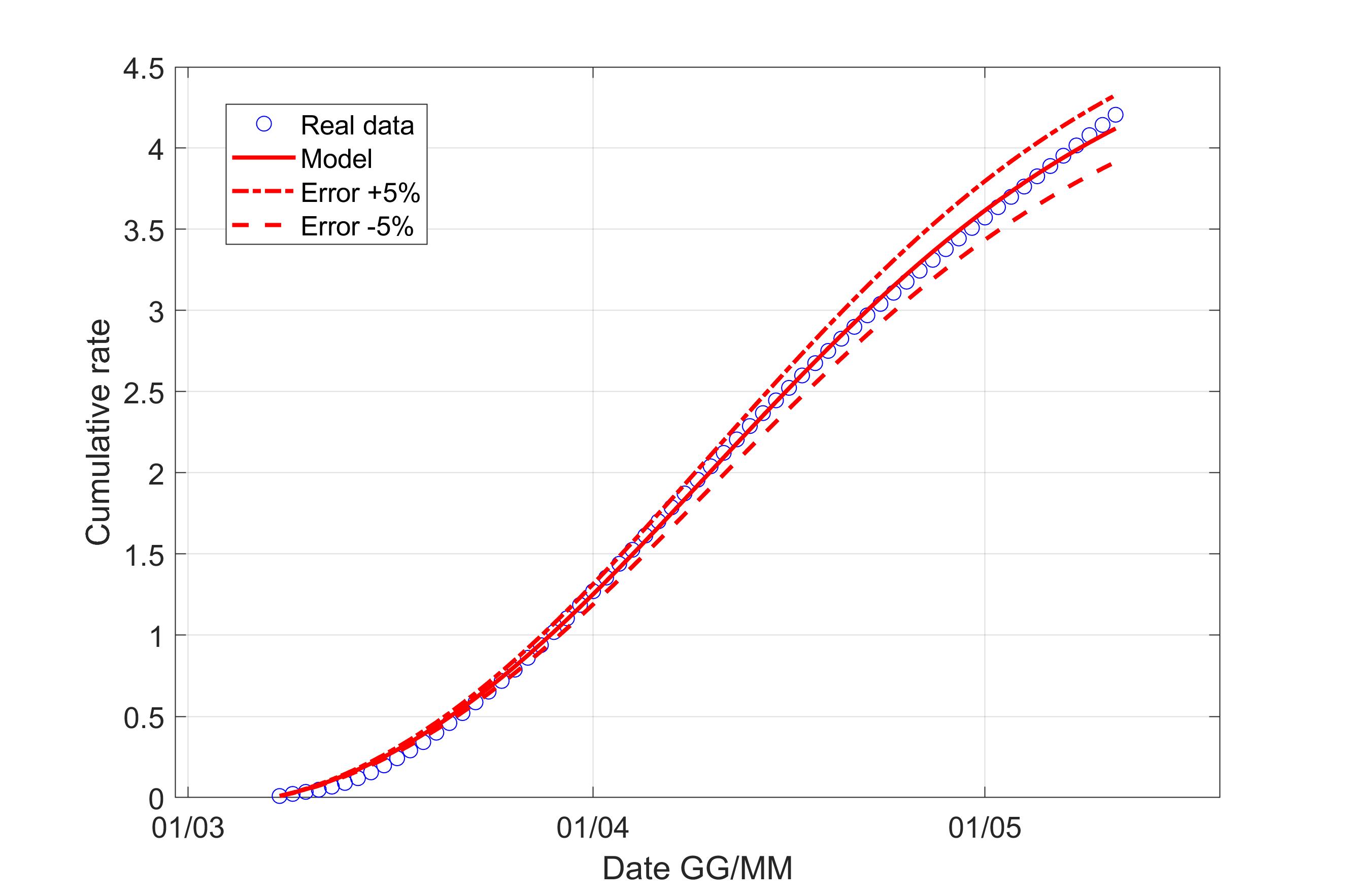}
\caption{The scenario of Germany with the derivative weight.} 
\label{Scenario3}
\end{figure}
\begin{figure}[hbt!]
\center\includegraphics[width=0.45\textwidth]{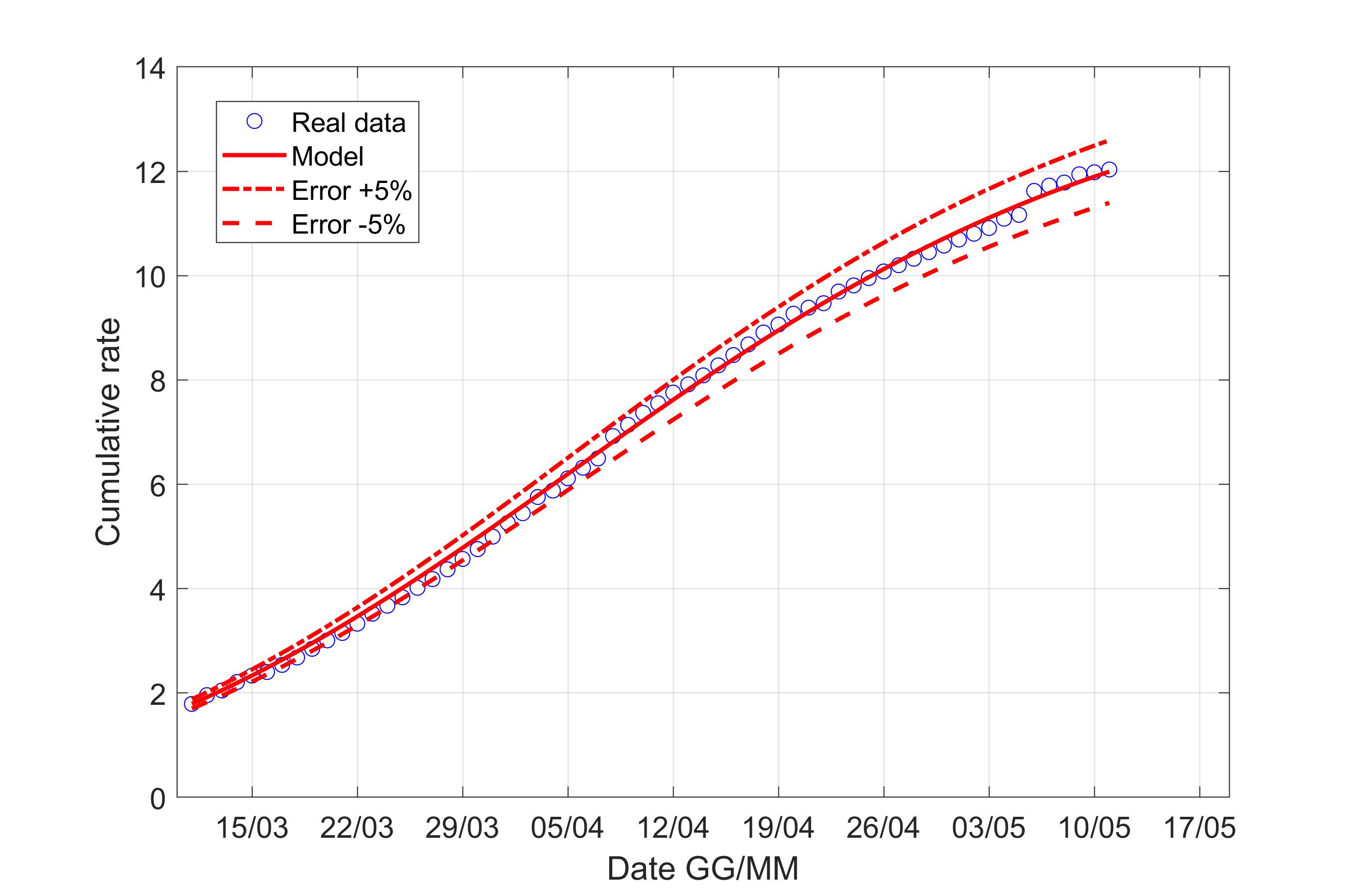}
\caption{The scenario of USA with the data weight.}
\label{Scenario4}
\end{figure}
A brief comment about these parameters: the values $r_{0}$ are similar to each other and lower than the Italian value, with and without Lombardia data. This is completely coherent with the various scenarios: the apparent lethality ( apparent CFR ) is approximately 4.5$\%$ in Germany and 6$\%$ in the United States, while in Italy is 14$\%$. The contribution of asymptomatic $A$ seems to be similar for all countries. \\
However the surprising result is that the trend of the virus is captured by the same differential equation, i.e. eq. (19), for all considered countries. This could mean that the virus always has the same type of behavior, but at different speed. Let's try to justify this idea: a different speed may depend on population density, work habits and the number of swabs at the beginning of the epidemic. About the last consideration we imagine to immediately carry out a large number of swabs: knowing as soon as possible the largest possible number of infected means limiting the contagion and therefore the propagation speed of virus.

\section{Conclusions}
We described the evolution of the Sars-Cov-2 in Italy in the time window February 24-May 5. To do this we have built a phenomenological growth model adapted on the data of Civil Protection. With respect to a classical SiR(D) model we did not consider the supsceptible population, because there are not medical evidences on which sample of the population can be ill. So we have considered three couple differential equations for Infected I(t), Deaths D(t) and Recovery R(t) with the a conservation law including the currently positive population P(t).\\
As the time delay between the onset of symptoms and healing $t_{r}$ days is a very oscillating parameter we introduced a sort of regressive function $f(t)$ to modelling better this delay. So we described two scenarios of the end of epidemic:
\begin{itemize}
\item $I(\mbox{end})=247471$, $D(\mbox{end})=35235$, close to July 10, for $f(t)$ linearly approximated,
\item $I(\mbox{end})=243766$, $D(\mbox{end})=34682$, close to June 20, for $f(t)$ in a quadratic approximation.
\end{itemize}
Obviously a linear approximation for f(t) leads to a slower recovery curve and therefore a small increase of infected.\\
In the second part of this manuscript we described the time evolution of the normalized data
\begin{equation}
I_{\mbox{norm}}(t_{i})=\frac{I(t_{i})}{S(t_{i})},
\end{equation}
that represents the number of infected normalized with the number of swabs $S(t_{i})$. \\
We have studied this parameter on four different scenarios:
\begin{itemize}
\item Italy,
\item data of Italy minus data of Lombardia ( about 37$\%$ of the Italian infected belong to the Lombardia region ),
\item USA,
\item Germany.
\end{itemize}
So we have found that all the evolutions are governed by the same generalized logistic equation \cite{MM}, suggesting an universal feature of the propagation of Sars-Cov-2 virus. In particular the value of the parameter $r_{0}$ is in descending order compatible with the respective Apparent CFR ( ACFR )
\begin{itemize}
\item for Italy $r_{0}=0.175$ and $ACFR=14\%$,
\item for Italy-Lombardia $r_{0}=0.143$ and $ACFR=11\%$,
\item for USA $r_{0}=0.082$ and $ACFR=6\%$,
\item Germany $r_{0}=0.069$ and $ACFR=4,5\%$.
\end{itemize}
Finally we suggest that the data $I_{\mbox{norm}}(t_{i})$ is the most important parameter to control the propagation of the virus for a new inauspicious propagation of this virus in the world, because, knowing its universal feature, we can forward know the number of infected preparing a relevant number of swabs.
\newpage
\begin{center} \textbf{Acknowledgement}\end{center}
We thank many colleagues for interesting discussions, in particular Andrea Marzolla and Domenico Seminara. We also thank Pierluigi Blanc, S.O.C. Infectious Diseases 1 Santa Maria Annunziata Hospital, for stimulating discussions on technical subjects on which we had no knowledge.

\end{document}